\documentclass[12pt]{article} 
\usepackage{epsfig}
\usepackage{rotating}

\def\vev#1{\left\langle #1\right\rangle}
\begin{document}

\begin{titlepage}

\title{ Maximal Neutrino Mixing from an\\ Attractive Infrared Fixed Point }

\author{
James Pantaleone$^1$, 
T. K. Kuo$^2$, 
Guo-Hong Wu$^3$ \\ \\ 
{\small {\it
$^1$Department of Physics, University of Alaska, Anchorage,
 Alaska 99508, 
\thanks{Email: tkkuo@physics.purdue.edu, jim@neutrino.phys.uaa.alaska.edu,
gwu@darkwing.uoregon.edu}}}\\
{\small {\it 
$^2$Department of Physics, Purdue University, West Lafayette, IN 47907
}}\\
{\small {\it $^3$Institute of Theoretical Science, University of Oregon,
Eugene, OR 97403}}\\ 
}

\date{\small (August, 2001)}   
\maketitle

\begin{abstract}
In the Standard Model (and MSSM), renormalization effects on neutrino mixing
are generally very small and the attractive fixed points are 
at vanishing neutrino mixing.
However for multi-higgs extensions of the Standard Model, 
renormalization effects on neutrino mixing can be large
and nontrivial fixed points are possible.
Here we examine a simple two-higgs model.
For two flavors, maximal mixing is an attractive 
infrared fixed point.  For three flavors, the neutrino mass matrix 
evolves towards large off-diagonal elements at low energies.  
The experimentally suggested bimaximal neutrino mixing pattern 
is one possible attractive infrared fixed point.

\end{abstract}

\vfill \end{titlepage}

\section{Introduction}

Recent experiments \cite{atmos,solar} have revealed important features 
of the neutrino mass matrix. It is now rather well-established that the neutrino
masses are tiny, and that at least some of the mixing angles are 
large, or even maximal. Considerable effort has been devoted to a 
theoretical understanding of these features.

The small value of neutrino masses is a natural feature of theories
with new physics entering at a high energy scale, such as Grand Unified Theories (GUT).
In such theories, neutrino masses enter 
the low energy effective Lagrangian as a dimension five operator \cite{dim5} involving
two fermion fields and two higgs fields. 
Thus the neutrino masses are suppressed from the charged fermion masses
by a factor of the ratio of the weak scale to the scale of new physics.
However these dimensional arguments do not explain why the neutrino mixing angles should be large.

Radiative corrections renormalize all terms in the Lagrangian, including the neutrino mass.
These effects may provide a dynamical explanation for the value of the top quark and Higgs masses
(for a recent review, see e.g. \cite{review}).
For neutrinos, these corrections were worked out some time ago \cite{BLP,CP} and have been studied
extensively of late for the Standard Model (SM) and the minimal supersymmetric standard model (MSSM)
(see e.g. \cite{Casas,other,KPW}).  In these models renormalization effects on neutrino mixing
are proportional to the charged lepton masses and hence are relatively small.
Also, maximal neutrino mixing is a saddle point in the SM and MSSM \cite{KPW}, with the infrared 
attractive fixed point corresponding to vanishing neutrino mixing.
Thus renormalization effects in the SM and MSSM can not readily 
explain the observed large neutrino mixing.

Here we shall study neutrino mixing in an extension of the Standard Model 
to two Higgs doublet fields, $\Phi_1$ and $\Phi_2$.
It is customary to assume that the two Higg's transform differently under a
discrete \cite{dissym} or continuous (e.g. \cite{PQ}) symmetry chosen 
so that each type of charged fermion couples to only one Higgs doublet.  
This insures that there are no flavor changing neutral Higgs couplings in the dimension four terms.
However this symmetry does not necessarily mean that there is only one type of 
dimension five neutrino mass term.  
In general, there are four ways to combine two Higgs fields and two neutrino fields \cite{BLP}.
Of the four dimension five operators, one involves two $\Phi_1$ fields, 
one involves two $\Phi_2$ fields,
and the other two depend on both fields.
The symmetry used to eliminate flavor changing neutral currents among the charged leptons
may be chosen such that almost any desired combination of these operators are allowed.
Here we choose to study the case where only the latter two operators are relevant:
\begin{equation}
{\cal L}_{\nu\nu} = {1 \over 2} \kappa_{mn} \bar{l^c}^m_{Li} l^n_{Lj}
\Phi_1^k \Phi_2^l
(\epsilon_{ik} \epsilon_{jl} - {1 \over 2} \epsilon_{ij} \epsilon_{kl} )
+ {1\over 2} \xi_{mn} \bar{l^c}^m_{Li} l^n_{Lj} \Phi_1^k \Phi_2^l
\epsilon_{ij} \epsilon_{kl}
+ h.c.
\end{equation}
where $l_{L}$ is the left-handed lepton doublet, $m$ and $n$ the generation indices
and $i, j, k, l$ are SU(2) indices.
\(\kappa_{mn}\) is symmetric under interchange of the generation indices \(m\) and \(n\),
while \(\xi_{mn}\) is antisymmetric.
When the Higgs' fields acquire vacuum expectation values \(\vev{\Phi^0_i} = v_i/\sqrt{2} \)
the neutrino mass term is  
\begin{equation}
M_\nu = - {1 \over 2} \kappa v_1 v_2 = - {1 \over 4} \kappa v^2 \sin 2 \beta.
\end{equation}
where $v^2 = v_1^2+v_2^2 = (246)^2$ GeV and $\tan \beta = v_2/v_1$.
Because the neutrino mass term does not explicitly depend on $\xi$, 
this combination of operators is relatively simple to study.

Renormalization mixes the two operators so both 
must be simultaneously evolved to study neutrino mixing.
In the basis where the charged lepton mass matrix is diagonal, 
the evolution equations are \cite{BLP} 
\begin{eqnarray}
{d \kappa_{\alpha\beta} \over d t}
& = & [ - 3 g_2^2 + 2 \lambda_3 + 2 \lambda_4 + S ] \kappa_{\alpha\beta}
- {1\over 2} (y_\beta^2 + y_\alpha^2) \kappa _{\alpha\beta}
\nonumber \\
& & + 2 (y_\beta^2 - y_\alpha^2) \xi_{\alpha\beta} ],
\nonumber \\
{d \xi_{\alpha\beta} \over d t}
& = & [ - 9 g_2^2 + 2 \lambda_3 - 2 \lambda_4 + S ] \xi_{\alpha\beta}
+ {3\over 2} (y_\beta^2 + y_\alpha^2) \xi_{\alpha\beta}
\nonumber \\
& & + {3\over 2} (y_\beta^2 - y_\alpha^2) \kappa_{\alpha\beta} ] .
\label{dk}
\end{eqnarray}
Here $t={1\over 16\pi^2} \ln \mu$, \(g_2\) is the SU(2)
gauge coupling constant,
and $y_\alpha = \sqrt{2} m_\alpha / v_1 = \sqrt{2} m_\alpha / v \cos \beta$ 
is the Yukawa coupling for charged leptons of type $\alpha$ (e, $\mu$ or $\tau$).
The quantity $S$ is defined as
\begin{equation}
S = Tr[ 3 Y_u^\dagger Y_u + 3 Y_d^\dagger Y_d + Y_e^\dagger Y_e ]  .
\label{S}
\end{equation}
where $Y_\alpha$ is the $3 \times 3$ Yukawa coupling matrix for charged fermions of type $\alpha$.
The $\lambda_i$'s are parameters from the higgs potential
\begin{eqnarray}
{\cal L}_{2H} & = & - { \lambda_1 \over 2} ( \Phi_1^\dagger \Phi_1 )^2
- { \lambda_2 \over 2} ( \Phi_2^\dagger \Phi_2 )^2
- \lambda_3 (\Phi_1^\dagger \Phi_1) (\Phi_2^\dagger \Phi_2)
\nonumber \\
& & - \lambda_4 (\Phi_1^\dagger \Phi_2) (\Phi_2^\dagger \Phi_1)
- [ {\lambda_5 \over 2} (\Phi_1^\dagger \Phi_2)^2 + h.c. ]
\end{eqnarray}
These equations assume $\Phi_1$ couples to the charged leptons,
but do not depend on which Higgs field couples to which 
type of charged quark.

In the limit that $\xi \rightarrow 0$, the evolution equation for $\kappa$ 
decouples and becomes similar to the SM case.   
We concentrate here on studying the new effects on neutrino mixing introduced 
by the $\xi$ parameter.
To simplify the analysis we assume that all parameters are real.

\section{Two neutrino flavors.}

The evolution of the scale of the neutrino mass matrix can be separated out
from the evolution of the dimensionless parameters.  
For dimensionless physical parameters in the two-flavor 
approximation we use the mixing angle $\theta$
\begin{equation}
{\rm tan}2\theta = {{2 \kappa_{\mu\tau}} \over {\kappa_{\tau\tau} - \kappa_{\mu\mu}}},
\end{equation}
a ratio of neutrino mass eigenvalues, $m_i$, defined as
\begin{eqnarray}
z & = & {{m_2 + m_1} \over {m_2 - m_1} } \nonumber \\
  & = & {{\kappa_{\tau\tau} + \kappa_{\mu\mu}} \over 
{\sqrt{ ({\kappa_{\tau\tau} - \kappa_{\mu\mu}})^2 + (2 \kappa_{\mu\tau})^2  }}} \ \,
\end{eqnarray}
and the ratio of the upper off-diagonal elements of the two dimension five terms
\begin{equation}
\eta = {\xi_{\mu\tau} \over \kappa_{\mu\tau}}.
\end{equation}
Since the diagonal elements of $\xi$ vanish, this parameterization is complete. 
The particular form of $z$ is chosen so that the fixed points are finite and simple \cite{KPW}.
In this parameterization,
\begin{equation}
M = { m_2 - m_1 \over 2} 
 \left(\begin{array}{cc}
z - \cos 2 \theta & \sin 2 \theta \\
\sin 2 \theta & z + \cos 2 \theta \\
\end{array}
\right)
\label{Mmatrix}
\end{equation}

After a little algebra, one finds that the 
evolution equations for the dimensionless parameters are
\begin{eqnarray}
{{d \eta} \over {d t}} & = & {3 \over 2} y^2 + C \eta - 2 y^2 \eta^2  \nonumber \\
{{d z} \over {d t}} & = &  y^2 ( {1 \over 2} 
\cos 2 \theta (z^2 -1)  - 2 z \eta \sin^2 2 \theta  ) \nonumber  \\
{{d \theta} \over {d t}} & = &  y^2 \sin 2 \theta ( {1 \over 4} z  + \cos 2 \theta \eta  )
\label{evo}
\end{eqnarray}
where 
\begin{eqnarray}
C    &=& - 6 g_2^2 - 4 \lambda_4 + 2 ( y_\tau^2 + y_\mu^2 ) \nonumber \\
y^2  &=& y_\tau^2 - y_\mu^2 
\end{eqnarray}

These equations and the dimension five operators 
possess various symmetries.  For example, under the transformation
\begin{equation}
\theta \rightarrow {\pi \over 2} - \theta \\
\label{st}
\end{equation}
accompanied by an interchange of the mass eigenvalues
\begin{equation}
z \rightarrow  - z \ \  ( m_2 \leftrightarrow m_1 ) \\
\label{sz}
\end{equation}
the diagonal elements of the neutrino mass matrix, Eq.~(\ref{Mmatrix}), 
are invariant while the off-diagonal elements change sign.
However the overall sign of the off-diagonal elements
is not a physical observable since it may be absorbed
in the unphysical phases with a redefinition of the neutrino wave function
$(\nu_\mu , \nu_\tau) \rightarrow (\nu_\mu ,- \nu_\tau)$.
This transformation changes the sign of $\xi_{\mu\tau}$ and $\kappa_{\mu\tau}$ simultaneously,
so $\eta$ is invariant.
Thus the neutrino mass matrix and the RGE evolution equations respect this symmetry.

The running of $\eta$ is independent of the neutrino physical parameters $z$ and $\theta$.
Thus we shall use $\eta$ to characterize the possible motion of $z$ and $\theta$.
The fixed points of $z$ and $\theta$
are obtained by setting their derivatives in Eq. (\ref{evo}) equal to zero and solving.
The stabilities of each fixed point are obtained by finding the eigenvalues of the
Jacobian at the fixed point (see e.g. \cite{strogatz}).  
The results of this analysis for decreasing t (i.e. approaching the infrared) are in the Table 
(for increasing t the stabilities are reversed).
As the Table shows, the location of the fixed points are independent of $\eta$,
but their stability does depend on $\eta$.
Of the five fixed points, only three are physically distinct since the symmetry of 
Eqs.~(\ref{st}) and (\ref{sz}) 
maps two into two others.
The fixed points at  $\theta = 0 , z = +1$ and $\theta = \pi/2 , z = -1$ 
correspond to no mixing and a 
massless muon-neutrino.
The fixed points at $\theta = 0 , z = -1$ and 
$\theta = \pi/2 , z = +1$ correspond
to no mixing and a massless tau-neutrino.
The fixed point at $\theta = \pi/4 , z = 0$ corresponds to maximal mixing with equal magnitude 
but opposite sign mass eigenvalues.
Note that the SM corresponds to $\eta \rightarrow 0$ for which large 
mixing is a saddle-node \cite{KPW}.

To determine which of the possibilities in the Table is realized, 
we must determine the behavior of $\eta$.
Setting ${d \eta \over d t } = 0$ in Eq. (\ref{evo}) yields the fixed points of $\eta$, 
$\eta_{\pm}^*$, as
\begin{eqnarray}
\eta_{\pm}^* &=& {1 \over 2} ( b \pm \sqrt{b^2 + 3} )  \nonumber \\
b &=&  {C \over {2 y^2}}  =  {{- 6 g_2^2 - 4 \lambda_4 + 2 ( y_\tau^2 + y_\mu^2 )  } 
\over {2 ( y_\tau^2 - y_\mu^2 ) }}  .
\label{afp}
\end{eqnarray}
Thus $\eta$ evolution has two fixed points---one positive and one negative.
The stability of the fixed points is easily found by plotting $-{{d \eta} \over {d t}}$
versus $\eta$, as in Fig. (1).  The minus sign is included in the derivative to give us
the stability for t {\it decreasing }.
We see that the positive fixed point, $\eta_+^*$, is a repellor while the negative fixed point, 
$\eta_-^*$, is an attractor.
This is true, regardless of the value of the parameters $C$ or $y^2$.
If the initial (high energy) value satisfies $\eta < \eta_+^*$, 
then $\eta$ evolves directly to the fixed point.
If the initial (high energy) value satisfies $\eta > \eta_+^*$ 
then $\eta$ initially evolves toward positive infinity.  
However note that evolution through $\eta = + \infty$ to the 
attractive fixed point at $\eta_-^*$ 
is possible when the denominator of $\eta$, $\kappa_{\mu\tau}$, evolves through 0.

From the Table, the attractive fixed point value of $\eta$ will cause 
maximal neutrino mixing to be an attractive fixed point when
\begin{equation}
\eta_-^* < - {1 \over 4}
\label{1o4}
\end{equation}
This corresponds to 
\begin{equation}
- \cos^2 \beta \left( {3 g_2^2 + 2 \lambda_4 \over y^2_{\rm SM}} \right) < 1.75
\label{1o4b}
\end{equation}
where $y^2 = y^2_{\rm SM} / \cos^2 \beta $ and $y^2_{\rm SM}$ is the
Standard Model value for the squared charged lepton Yukawa coupling differences
($y^2_{\rm SM} = 1.0 \times 10^{-4}$ for $y_\tau^2 - y_\mu^2$'s and 
$y^2_{\rm SM} = 3.5 \times 10^{-7}$ for $y_\mu^2 - y_e^2$).
Our lack of knowledge of the Higg's parameters $\tan \beta$ and $\lambda_4$ (especially
its sign) makes it
impossible to definitively evaluate this condition. 
However it appears that maximal neutrino mixing 
is an attractive fixed point for an extremely wide range of parameters.

The neutrino mixing and mass evolution for the case of negative $\eta$ is graphically displayed
in Fig. (2).  There the fixed points and the direction field are plotted for {\it decreasing} t.
Starting from some initial point specified by the high energy theory,
the evolution of $z$ and $\theta$ follow a trajectory on this graph and the
plotted unit arrows show the directions tangent to this trajectory.
The vector field plot does not explicitly depend on $t$ or $y^2$ but does depend on the
value of $\eta$.  
However for $|\eta| >> 1$ the explicit $\eta$ dependence also 
cancels out of the vector field plot.
Here we use a fixed point value $\eta_-^* = -900$
which corresponds to $g_2^2 = 0.4$, $\lambda_4 = 1$ and $y^2 = 3.5 \times 10^{-3}$. 
Except at the extreme upper and lower edges of the plot, the evolution is towards the 
large mixing fixed point.

In the Standard Model, the small size of $y^2$ tends to suppress the
neutrino parameter evolution.   To compensate for this small factor,
previous analyses \cite{BLP,CP,Casas,other} have usually focused on the possibility of a
large neutrino mass degeneracy at high energies, i.e. $|z| \ge 1 / y^2$.
Let us briefly consider the evolutionary behavior of large $|z|$.
For large $|z|$, $z$ changes much faster than $\theta$ does.
For $|z| > > |\eta|$, 
the evolution equations, Eq. (\ref{evo}), are the same as those of the Standard Model and
the approximate invariant is \cite{KPW}
\begin{equation}
{\sin^2 2 \theta \over z^2 - 1 } \approx \zeta
\end{equation}
where $\zeta$ is a constant, independent of $t$ and $y^2$.
Qualitatively, the running is away from the Standard Model repulsive fixed point and toward the
Standard Model attractive fixed point (see Table).
For small initial (high energy) $\theta$ values,
running causes $z>0$ and $\theta$ to decrease 
until $z \approx \eta$ when $\eta$ effects become relevant and then drive the mixing 
towards maximal.
For large initial $\theta$ values ($\theta$ near $\pi/2$),
running causes $z>0$ to initially increase and $\theta$ to initially decrease, 
until $\theta$ evolves through maximal mixing to $\theta < \pi/4$ where $z$ starts to decrease.  
When $z$ has decreased enough such that $|z| \sim |\eta|$ 
then $\eta$ effects become relevant and drive the mixing back toward maximal.
Thus it is only when $|\eta| \ge |z|$ that the attractive fixed point nature of 
maximal mixing is apparent.
Sufficient evolution will eventually produce $|\eta| >> |z|$.

For large $|\eta|$, the $z$ dependence of $ d \theta \over d t$ in Eq. (\ref{evo}) disappears
and evolution of the mixing angle simplifies considerably.
Then the renormalization equations may be solved to give
\begin{equation}
\tan 2 \theta_{\rm min} \approx \tan 2 \theta_{\rm max} 
{\rm Exp}{[ 2 \int_{t_{max}}^{t_{min}} dt y^2 \eta ]}
\label{thtan}
\end{equation}
which describes the evolution of the mixing angle.
Here $t_{max}$ and $t_{min}$ are the high energy and low energy values of $t$.
This equation clearly shows that running does not change the sign of $\tan 2 \theta$,
but just increases its magnitude, i.e. $\theta \rightarrow \pi/4$.
From Eq. (\ref{thtan}), the $z$ evolution can be obtained using the large $\eta$ 
approximation invariant
\begin{equation}
{z \over \cos 2 \theta } \approx \chi
\end{equation}
where $\chi$ is a constant, independent of $\eta$, and so only slowly varying with $t$ and $y^2$.

Using Eq. (\ref{thtan}) we can find the approximate conditions for significant evolution toward the
maximal mixing fixed point.
A change in $\tan 2 \theta$  greater than a factor of 10 requires
\begin{equation}
6.1 < \vev{ y^2 ( - \eta ) }
\label{tcon}
\end{equation}
where we have taken $\mu_{\rm min} = 10^2 {\rm GeV}$, $\mu_{\rm max} = 10^{15} {\rm GeV}$ and 
the $\vev{}$ denotes a value averaged over the running between these points.
This condition may be satisfied if $|\eta|$ is large and/or $\tan \beta$ is large, 
since $y^2 = y^2_{SM}/\cos^2 \beta$.
The validity of perturbation theory requires $y^2 < 4 \pi$.
It is interesting to note that if $y^2$ is small,
i.e. $| 6 g_2^2 + 4 \lambda_4| >> y^2$ and we take $\eta \approx \eta_-^*$, 
the attractive fixed point value (Eq. (\ref{afp})), 
then the dependence on the charged lepton Yukawa coupling disappears from Eq. (\ref{tcon}) to give
\begin{equation}
\vev{\lambda_4} > {1 \over 4} \left[ {\ln (10) \over t_{\rm max}-t_{\rm min}} - 
6 \vev{g_2^2} \right] \approx 2.4
\label{tcon2}
\end{equation}
where we have used $g_2^2 \approx 0.4$. 
The validity of perturbation theory requires $\lambda_4 < 8 \pi$.

The evolution of $\eta$ may be solved for approximately.
If we neglect the evolution of $y^2$ and $C$, then $d \eta \over d t$ in Eqs. (\ref{evo})
may be integrated to give
\begin{equation}
{ 
(\eta_{\min} - \eta_+^*) (\eta_{\max} - \eta_-^*)
\over 
(\eta_{\max} - \eta_+^*) (\eta_{\min} - \eta_-^*) 
} 
= {\rm Exp}[ 2 y^2 (\eta_+^* - \eta_-^*) (t_{max} - t_{min}) ]
\label{econ}
\end{equation}
Using this equation, we may estimate the condition for evolution to produce
large values of $\eta$.  
Let us take $\eta_{max} \approx -1$, $\eta_{min} \approx {\eta_-^* \over 2} < < -1$.
Then Eq. (\ref{econ}) reduces to
\begin{equation}
| \eta_-^* | < {\rm Exp}[ 2 y^2 | \eta_-^* | (t_{max} - t_{min}) ]  .
\label{econ2}
\end{equation}
If $y^2$ is small, i.e. $| 6 g_2^2 + 4 \lambda_4| >> y^2$, then Eq. (\ref{econ2}) reduces to
\begin{equation}
 \ln \left[ {1.2 + 2 \lambda_4 \over y^2 } \right] < [ 0.46 + 0.76 \lambda_4 ]
\label{econ3}
\end{equation}
Unlike Eq. (\ref{tcon2}), this conditions still weakly depends on $y^2$. 
The smaller $y^2$ is, the larger the magnitude of the attractive fixed point value $\eta_-^*$,
and so more running is required which in turn requires a larger $\lambda_4$.

In general, the amount the mixing angle evolves depends on two parameters that are not 
well known, $\lambda_4$ and $\tan \beta$.
A complete analysis of all the constraints on these parameters, and how they relate to 
neutrino mixing angle evolution, is beyond the scope of the present paper.
However we note that studies of the evolution of the two-Higgs 
potential parameters (see e.g. \cite{2hrge})
suggest that the $\lambda_i$'s are not generally driven to a particular fixed point, 
but that a range of values are consistent with experimental and theoretical constraints.
Constraints on two-Higgs models also follow from studies of flavor changing neutral currents.  
In particular, the flavor changing neutral current decay b $\rightarrow$ s + $\gamma$ 
tightly constrains two-Higgs models where a different Higgs 
couples to the two types of quarks (type-II)
(see e.g. \cite{FCNC}).  
However this constraint can be avoided by coupling $\Phi_2$ to both 
quark types and $\Phi_1$ to the charged leptons.
Then large $\tan \beta$ corresponds to large charged lepton Yukawa couplings,
but approximately Standard Model values of the quark Yukawa couplings.
Large values of $y^2$ would help produce large mixing angle evolution, 
but they also tend to enhance lepton universality violations.
However a recent analysis of W and Z decay data \cite{LLT} finds very weak limits on
two-Higgs' models from lepton universality, 
with values of $\tan \beta$ up to and exceeding 100 allowed
(this corresponds to $y_\tau$ exceeding 1).
Thus large running of the neutrino mixing angle appears to be compatible with
present constraints on two-Higgs models.

\section{Three neutrino flavors.}

The evolution of the physical mixing angles and mass ratios 
is quite complicated for three flavors.  
Here we do not give explicit formulae for their evolution but
instead focus on the texture of the mass matrix produced by evolution.
Then we plot renormalization effects obtained numerically 
for one particular scenario.

The general nature of renormalization effects 
on the neutrino mass matrix in this model
can be found by examining the relative running of diagonal to off-diagonal elements
\begin{equation}
{d \over d t} \left( { \kappa_{\beta \beta} \over \kappa_{\alpha \beta}} \right)
= -2 ( y_\beta^2 - y_\alpha^2 ) \left[ \eta_{\alpha \beta} + {1 \over 4} \right] 
\left( {\kappa_{\beta \beta} \over \kappa_{\alpha \beta} } \right)  .
\label{dbbab}
\end{equation}
Here $\eta_{\alpha \beta}$ is defined just as it was for two flavors,
\begin{equation}
\eta_{\alpha \beta} = {\xi_{\alpha \beta} \over \kappa_{\alpha \beta}} .
\end{equation}
$\eta_{\alpha \beta}$ is  antisymmetric so for three flavors there are now 
three independent $\eta$'s; $\eta_{e\mu}$, $\eta_{e\tau}$ and $\eta_{\mu\tau}$.
For two-flavor, this equation provides another way to understand the Table.  
For $\eta_{\alpha \beta} < - 1/4$ for $\beta > \alpha$, the coefficient
of $( { \kappa_{\beta \beta} / \kappa_{\alpha \beta}} )$
on the right-hand side is positive for $\beta > \alpha$, and also for $\alpha > \beta$ since
$\eta_{\alpha \beta}$ is antisymmetric.  
Thus as $t$ decreases the two diagonal elements decrease with respect to the off-diagonal elements.
This leads to maximal mixing, in agreement with the Table.
Conversely for $ 1/4 < \eta_{\alpha \beta} $ for $\beta > \alpha$, 
both diagonal elements increase with respect to
the off-diagonal elements, 
which drives the mixing away from maximal mixing and towards vanishing mixing.
Standard Model like behaviour occurs for $ - 1/4 < \eta_{\alpha \beta} < 1/4$.
Then the cofficient of $( { \kappa_{\beta \beta} / \kappa_{\alpha \beta}} )$
on the right-hand side is positive for $\beta > \alpha$ but negative for $\alpha > \beta$.
Thus as infrared energies are approached the lower (upper) 
diagonal element increases (decreases) with respect
to the off-diagonal term.  Running can go through the point where the diagonal elements are equal, 
and maximal mixing occurs, and proceed on towards vanishing mixing.
Thus the texture of the $\kappa$ matrix depends on the behaviour of the $\eta_{\alpha \beta}$'s.

The dynamics of the $\eta_{\alpha \beta}$'s follows from Eqs. (\ref{dk}).
There we see that $\xi_{\alpha \beta}$ and $\kappa_{\alpha \beta}$
only mix with each other so, aside from different Yukawa couplings,
the evolution equation for each
three-flavor $\eta_{\alpha \beta}$ is identical in form to the two-flavor
evolution equation for $\eta$ given in Eqs. (\ref{evo}).  
Thus the general dynamics of the $\eta_{\alpha \beta}$ for $\beta > \alpha$
are identical to what has 
been discussed previously and plotted in Fig. (1).  
They all evolve toward attractive fixed points, $\eta_{\alpha \beta}^*$, 
given by the negative root of Eq. (\ref{afp}).  
Thus evolution produces $\eta_{\alpha \beta}$'s which, for $\beta > \alpha$, are negative 
and readily less than $- 1/4$ (see Table and Eqs. (\ref{1o4}) and (\ref{1o4b})).
Referring back to Eq. (\ref{dbbab}), this implies that evolution toward low energies
decreases the diagonal elements relative to the off-diagonal elements of the neutrino mass matrix.
Then in the infrared limit the neutrino mass matrix approaches the form
\begin{equation}
\kappa_{\alpha \beta} \rightarrow 
\left[
\begin{array}{ccc}
0 & \xi_{e\mu}/\eta_{e \mu}^* &  \xi_{e\tau}/\eta_{e \tau}^* \\
\xi_{e\mu}/\eta_{e \mu}^* &  0 & \xi_{\mu\tau}/\eta_{\mu \tau}^* \\
\xi_{e\tau}/\eta_{e\tau}^* &  \xi_{\mu\tau}/\eta_{\mu \tau}^* & 0
\end{array}
\right] .
\label{zee}
\end{equation}
Here the $\eta^*$'s are the attractive fixed point values and 
$\xi$'s are their values at low energies.

For comparison purposes, let's consider the relative running of two off-diagonal elements
\begin{equation}
{d \over d t} \left( { \kappa_{e \mu} \over \kappa_{e \tau}} \right)
= \left[ 
{1 \over 2} ( y_\tau^2 - y_\mu^2 ) + 
2( y_\mu^2 - y_e^2 )  \eta_{e \mu} - 
2( y_\tau^2 - y_e^2 )  \eta_{e \tau}  
 \right] 
\left( {\kappa_{e \mu} \over \kappa_{e \tau} } \right)
\label{dod}
\end{equation}
This equation shows that a cancellation occurs between the two terms proportional to $\eta$'s.
Indeed, if we substitute in the fixed point value of the $\eta$'s for small Yukawa
couplings, all large terms cancel out and the running is not significant.
Thus the running of the off-diagonal terms is generally not a dominant process.

The form of the neutrino mass matrix in Eq. (\ref{zee}) is the same as that produced
in the popular Zee model \cite{Zee}.
However note that the Zee texture is never exactly obtained from renormalization, 
i.e. the diagonal elements never completely vanish.
How closely the Zee texture is approached depends on the amount of evolution, 
the size of the initial (high energy) parameters 
and the values of the two-higgs parameters $\tan \beta$ and $\lambda_4$.
But this general texture is interesting because 
it can describe the experimentally desired bimaximal neutrino mixing \cite{bimax}
when $\kappa_{e\mu} \approx \kappa_{e\tau} >> \kappa_{\mu\tau}$.
This particular relationship among the off-diagonal parameters
is not caused by running, as Eq. (\ref{dod}) shows.  
However it may be realized if the $\xi_{\alpha\beta}$'s at the high energy scale
have the appropriate hierarchy of values.

The possible size and nature of running are demonstrated in Figs. (3) and (4).
In these calculations, 400 different neutrino mass matrices were generated
by choosing the elements of $\kappa_{\alpha\beta}$  
at the high energy scale to be real, 
random numbers evenly distributed between -1 and 1.
The $\xi_{\alpha\beta}$'s at the high energy scale were chosen to be
$\xi_{e\mu} =2600$, $\xi_{e\tau} =10$ and $\xi_{\mu\tau} = 0$.
The parameters $\xi_{e\mu}$ and $\xi_{e\tau}$ were chosen to be larger 
than the $\kappa_{\alpha\beta}$'s at high energies to insure that 
$\kappa_{e\mu}$ and $\kappa_{e\tau}$ would be the largest matrix elements at low energies.
Additionally, these parameters were chosen because they satisfy 
$\xi_{e\mu} / \xi_{e\tau} \approx \eta_{e\mu}^* / \eta_{e\tau}^*$
(actually their ratio is a little less than this to allow for running
before the $\eta_{\alpha\beta}$'s reach their fixed points).
The other relevant parameters were taken to be $g_2^2 = 0.4$, $\lambda_4 = 1$,
$y_\tau = 1$ (which is equivalent to $\tan \beta = 100$), and
$[-3 g_2^2 + 2 \lambda_3 + 2 \lambda_4 + S] = 7.8$ 
(the last influences the overall neutrino mass scale evolution).
These parameters were not evolved but instead held
fixed because they mostly influence the overall amount of running so
their precise values are not particularly important, and because the $\lambda_i$'s and 
$\tan \beta$ are for the most part unknown.

Fig. (3) shows the mixing angles at the high and low energy scales.
The mixing angles are defined following the convention of ref. \cite{KP} as
\begin{equation}
\sin^2 2 \psi = {4 U_{\mu 3}^2 U_{\tau 3}^2 \over [ 1 - U_{e 3}^2 ]^2 }
\end{equation}
and
\begin{equation}
\sin^2 2 \omega = {4 U_{e 1}^2 U_{e 2}^2 \over [ 1 - U_{e 3}^2 ]^2 }
\end{equation}
where $U_{\alpha i}$ is the unitary mixing matrix.
The neutrino mass eigenvalues are ordered such that 
\begin{eqnarray}
| m_2^2 - m_1^2 | &<& | m_3^2 - m_1^2 | \\
| m_2^2 - m_1^2 | &<& | m_3^2 - m_2^2 |
\end{eqnarray}
so that solar neutrino oscillations are described by the $m_2^2 - m_1^2$ mass squared difference.
Thus $\psi$ is the mixing angle relevant for atmospheric "$\nu_\mu$-$\nu_\tau$" oscillations
and $\omega$ is the mixing angle relevant for solar neutrino oscillations.
Fig. (3) clearly demonstrates that renormalization from high to low
energies can simultaneously drive these angles to maximal.
The running effects on $U_{e3}$ are not shown,
however for the same parameters the relevant mixing quantity, 
$\sin^2 2 \phi = 4 U_{e3}^2 ( 1 - U_{e3}^2)$, 
is driven from a range of values between 0 and 1 to $\sin^2 2 \phi < 0.05$ at the low energy scale.
This is consistent with the experimental limits from CHOOZ \cite{CHOOZ}.
Thus running can drive the neutrino mass matrix to a bimaximal form, 
consistent with present experiments.

The approximate conditions for large mixing angle evolution were described in 
the two-flavor section in Eqs. (\ref{tcon}), (\ref{tcon2}) and Eqs. (\ref{econ2}), (\ref{econ3}).
In particular, these equations shows that large angle evolution results
if $\lambda_4$ is large.  Indeed, our numerical results remain essentially unchanged
if the charged lepton Yukawa couplings are decreased to be their
Standard Model values ($\tan \beta << 1$) and $\lambda_4$ is increased to $\lambda_4 > 16$.
This agrees with Eqs. (\ref{tcon2}) and (\ref{econ3}),
except that the numerical results are somewhat better than the analytical estimates
suggested by these equations because the numerical results used
initial values of $\xi_{e\mu}$ and $\xi_{e\tau}$ that were large in magnitude.
Large mixing angle evolution could be maintained for smaller 
values of these initial $\xi_{e\mu}$ and $\xi_{e\tau}$ by increasing the size of 
$\lambda_4$.

Fig. (4) shows how running affects neutrino mass squared 
differences.  The quantities $m_3^2 -m_1^2$ and $m_2^2 -m_1^2$ 
are each scaled by the largest neutrino mass squared and plotted against each other. 
From a distribution of values at the high energy scale,
evolution produces an inverted hierarchy of neutrino masses
with, roughly,  $m_3^2 << m_1^2 , m_2^2$.
This is consistent with the results found in studies of 
bimaximal mixing in the Zee model \cite{bimax}.

\section{Conclusions}

Neutrino masses enter into low energy physics as a dimension five operator,
regardless of the details of the high energy theory.
In multi-higgs extensions of the Standard Model,
there are generally more than one dimension five lepton-higgs operator.
Here we have worked in the two-higgs model, picked two of the dimension five operators
(one a symmetric mass term, the other an antisymmetric operator),
and examined how mixing angle evolution is changed from the Standard Model.
Renormalization mixes these different operators,
thus neutrino mixing depends on how the two operators evolve. 
Because the operators' evolution have different dependences on 
a higgs potential parameter, the mixing angle evolution depends on 
this parameter also.  This parameter can be much larger than the
tiny Standard Model charged lepton Yukawa couplings, 
so the mixing angle evolution can be quite large.
The size of angle evolution is also enhanced because
multi-higgs models have larger charged lepton Yukawa couplings.

For the model studied here, for a wide range of parameters, running tends to produce
a neutrino mass matrix whose diagonal elements are small compared to the off-diagonal elements.
For two neutrino flavors, this corresponds to maximal mixing as the 
attractive infrared fixed point.
For three neutrino flavors, the neutrino mass matrix resembles that
produced by the Zee model \cite{Zee}.
It is known that this model can describe the experimentally suggested bimaximal mixing.
Here we showed explicitly that bimaximal mixing can be an attractive infrared fixed point
if the elements of the asymmetric dimension five operator are large and heirarchical.

This model predicts that the neutrino mass spectrum is inverted,
with $m_3^2 << m_2^2, m_1^2$.
This possibility may be tested using matter effects \cite{invert},
either in long baseline experiments or in atmospheric neutrino observations. 
The scenario described here may be distinguished from the Zee model by
observing experimental processes such as double beta decay and CP violation.
These quantities are highly suppressed in the standard Zee model,
but are expected to only be moderately suppressed by renormalization effects.

\bigskip

 {\bf Acknowledgements:} \ 
T. K. K. and G.-H. W. are supported in part by DOE grant No.
DE-FG02-91ER40681 and No. DE-FG03-96ER40969, respectively.
J.P. is supported by NSF grant PHY-0070527.


\raggedbottom
\newpage
Table.  
The stability of the fixed points in our two-Higgs model for t {\it decreasing},
i.e.  the infrared limit. The notation is A = attractor, S = saddle, R = repellor.
\\
 
\begin{center}
\begin{tabular}{| c | c || c | c | c |} \hline
$\theta$ & $z$  & $\eta < - {1 \over 4} $ & $- {1 \over 4} < \eta < {1 \over 4}$ 
& ${1 \over 4} < \eta$  \\
\hline
\hline
$\pi/4$  & 0 & A & S & R   \\
\hline
0 & +1 & S & A & A   \\
${\pi \over 2}$  & -1 & S & A & A   \\
\hline
0              & -1 & R & R & S   \\
${\pi \over 2}$  & +1 & R & R & S   \\
\hline
\end{tabular}
\end{center}

\raggedbottom

\newpage

\begin{figure}[t]
\centerline{\psfig{figure=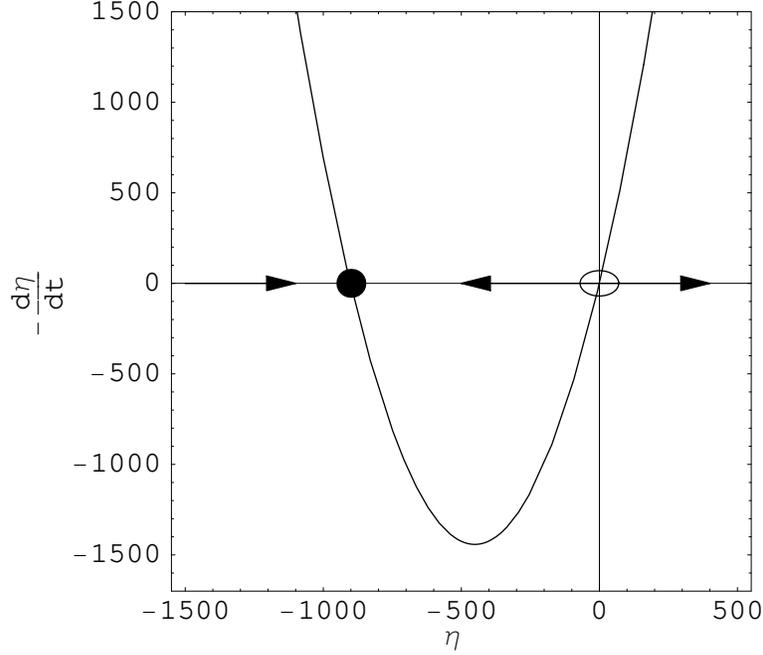,width=10.cm,angle=0}}
\caption[] {
Plot of $- {d \eta \over d t}$ versus $\eta$, the ratio of off-diagonal terms
between the two dimension five neutrino-higgs operators.
We have taken $g_2^2 = 0.4, y^2 = 3.5 \times 10^{-3}$ and $\lambda_4 = 1$.  
The solid circle and
open circle denote the attractive and repulsive infrared fixed points.
 \label{fig1} }
\end{figure}


\begin{figure}[b]
\centerline{\psfig{figure=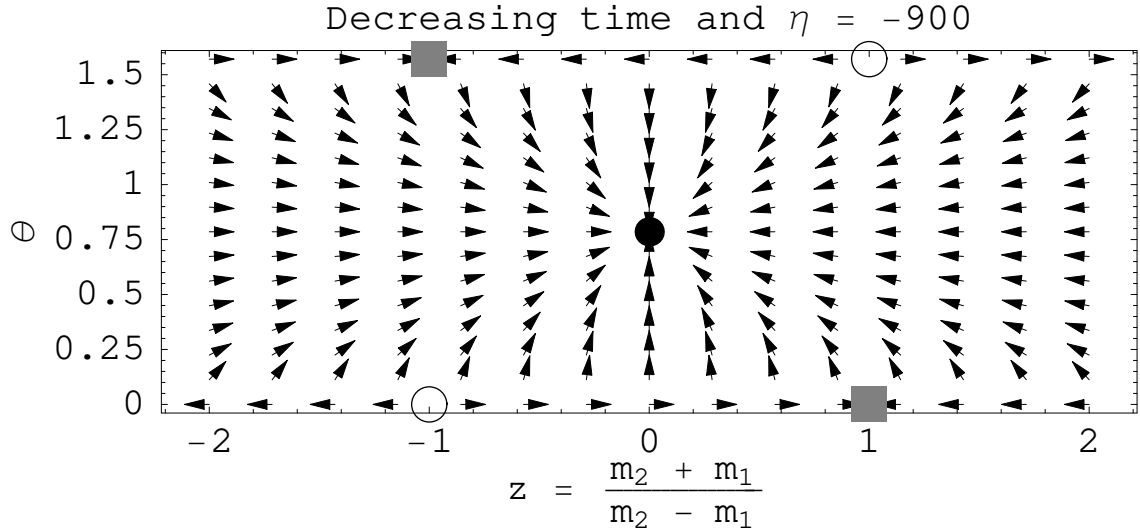,width=7.cm, angle=-90}}
\caption[] {
Direction field for running of the physical neutrino parameters.
Two neutrino flavors are assumed with $z$ being the ratio of
mass differences, $\theta$ the mixing angle and $\eta=-900$.
The different infrared fixed points are shown with solid circles, 
open circles and grey square denoting attractors, 
repellors and saddle point. 
 \label{fig2} }
\end{figure}

\newpage

\begin{figure}[htbp]
\centerline{\psfig{figure=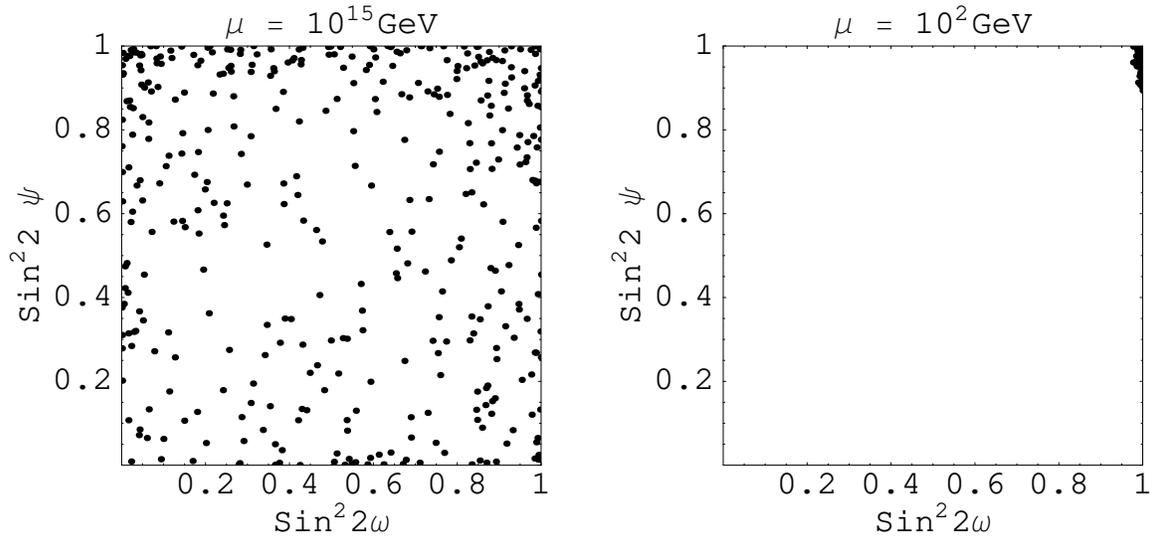,width=7.cm, angle=-90}}
\caption[] {
The plots show how evolution affects the solar 
$\sin^2 2 \omega$ and atmospheric $\sin^2 2 \psi$
neutrino mixing parameters.
Random neutrino mass matrices at
the high energy scale ($\mu = 10^{15}$ Gev) are evolved down to the low energy scale 
($\mu = 10^{2}$ Gev).
The fixed parameters are $g_2^2 = 0.4$, $\lambda_4 = 1$, $y_\tau = 1$, 
$\xi_{e\mu} =2600$, $\xi_{e\tau} =10$ 
and $\xi_{\mu\tau} = 0$.
\label{fig3} }
\end{figure}


\begin{figure}[b]
\centerline{\psfig{figure=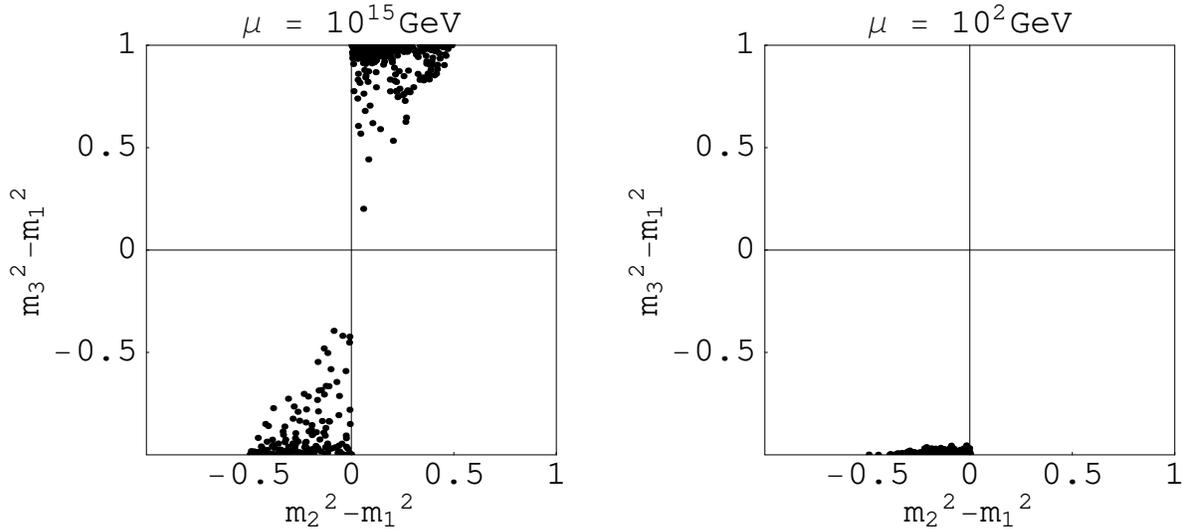,width=7.cm, angle=-90}}
\caption[] {
The plots show how evolution affects the neutrino mass squared differences.
Random neutrino mass matrices at
the high energy scale ($\mu = 10^{15}$ Gev) 
are evolved down to the low energy scale ($\mu = 10^{2}$ Gev).
The parameter choice is identical to that used in the previous figure.
The mass squared differences are scaled by the largest neutrino mass squared.
 \label{fig4} }
\end{figure}

\end{document}